\documentclass{article}

\usepackage{arxiv}

\usepackage[utf8]{inputenc} 
\usepackage[T1]{fontenc}    
\usepackage{hyperref}       
\usepackage{url}            
\usepackage{booktabs}       
\usepackage{amsfonts}       
\usepackage{nicefrac}       
\usepackage{microtype}      

\usepackage{mathtools}
\usepackage{environ}
\usepackage{xcolor}
\usepackage[tikz]{bclogo}
\usepackage{tikz}
\usetikzlibrary{calc}
\usepackage{amssymb}
\usepackage{todonotes}

\def\ket#1{|{#1}\rangle}
\def\bra#1{\langle{#1}|}
\def\Tr{\text{Tr}}
\def\no#1{{:\!#1\!:}}

\newcommand{\be}[0]{\begin{equation}}
\newcommand{\ee}[0]{\end{equation}}
\newcommand{\bea}{\begin{eqnarray}}
\newcommand{\eea}{\end{eqnarray}}
\newcommand{\nn}{\nonumber}

\title{On KdV characters in large $\mathbf{c}$ CFTs}

\author{Enrico M. Brehm\\
        Max Planck Institut f\"ur Gravitationsphysik,\\
    	Albert-Einstein-Institut, \\
	    Potsdam-Golm, D-14476, 
	    Germany.\\
        \texttt{brehm@aei.mpg.de} \\
   \And
        Diptarka Das\\
        Indian Institute of Technology, Kanpur,\\
	    Kanpur, Uttar Pradesh-208016,
	    India, and \\
        Max Planck Institut f\"ur Gravitationsphysik,\\
    	Albert-Einstein-Institut, \\
	    Potsdam-Golm, D-14476, 
	    Germany.\\
	      \texttt{didas@iitk.ac.in}
}

\begin{document}
\maketitle

\begin{abstract}
Two-dimensional conformal field theories with just Virasoro symmetry are endowed with integrable structure.
We review how to construct the integrable charges in a two-dimensional conformal field theory and how to relate them to the charges of quantum Sinh-Gordon theory when $c \geq25$. We then explicitly calculate the single charge characters in the large $c$ limit for all charges and thereby reveal how their degeneracies grow within one module. This, in particular, allows us to approximate the characters in the limit of small chemical potential, which source the respective charges. The latter give us insights into possible transformation properties of the characters. We also comment on the full generalized Gibbs ensemble and approximations to pure states.
\end{abstract}


\section{Introduction}

At the heart of the solvability of two-dimensional conformal field theories (CFTs) lies the infinite dimensional Virasoro algebra. This permits the construction of an infinite set of mutually commuting charges, first exhibited in \cite{sy} and also \cite{Eguchi:1989hs}. A deep connection between CFTs and integrable theories arises when quantising soliton equations (specifically, the Sine-Gordon equation). The charges themselves are constructed as integrals over (finite) polynomials of the energy momentom tensor and as such are polynomial functions of the Virasoro generators. Often one calls them the Korteweg-de-Vries (KdV) charges of the conformal field theory, $\hat{K}_n$, since they are classically related to the integrals of motion of the KdV equation (see \eqref{kdv-class}). Even though the infinite set of these charges in \cite{sy} are related to the Sine-Gordon theory for $c \leq 1$, there exists a proper analytic continuation to large $c$ CFTs that relate them to Sinh-Gordon theory, as we show below. 

All these charges commute with the Hamiltonian of the CFT, $[H_\text{CFT},\hat{K}_n] =0$. Thus they generate a rigid infinite dimensional abelian symmetry within the larger symmetry generated by the full enveloping Virasoro algebra. Therefore, the KdV charges are conserved quantities of the two-dimensional CFT. Due to their connection to the theory of integrable systems one can apply techniques thereof in the context of CFT. This was done in the series of remarkable papers \cite{blz,blz2,blz3}, where they map the problem of finding the exact eigenvalues of all these charges in the $c \leq 1$ CFTs to that of solving the pure quantum mechanical problem of a single particle in a specific potential. Moreover, under specific relevant perturbations, the integrability structure is retained. This has been utilized to find the exact S-matrix off-criticality \cite{2016JPhA...49F3006N,doi:10.1142/S0217751X8900176X} and thereby provides additional motivation to understand the charges in CFTs.

The KdV charges can also be used to define a generalized partition function, or the Generalized Gibbs Ensembles (GGE). When a critical system is in equilibrium with some environment such partition functions naturally arise.  If we do not specify the allowed interactions between the system and the environment then one needs to take all conserved quantities into account. Such a generic equilibrium state of the system with chemical potentials for every possible conserved quantity can be written 
\begin{equation}
\rho(\{q,y_i\}) = q^{H_\text{CFT}}\prod_{i\ge2}  y_i^{\hat{K}_i}\,,
\end{equation} 
where $q = e^{-\beta}$ with the inverse temperature $\beta$, and $y_i \equiv e^{-\mu_i}$ with the chemical potential $\mu_i$. The GGE itself is then given by
\begin{equation}
Z(\{q,y_i\}) = \Tr\left[ \rho(\{q,y_i\})\right] = \Tr_{[0]} \left[ \rho(\{q,y_i\})\right] + \sum_p \Tr_{[p]}  \left[ \rho(\{q,y_i\})\right]\,,
\end{equation}
where the first trace is over the full spectrum of the CFT, and $\Tr_{[p]}$ stands for the trace over a primary module $[p]$, with $[0]$ denoting the vacuum module. The subtraces over specific modules will naturally be called the GGE characters. 

Purely thermal states are obviously given by  $\rho(\{q,y_i =1\})$, i.e. all chemical potentials have to vanish. In two dimensional CFT, the thermal ensemble -- or partition function -- has a nice geometric description. If the theory is defined on a circle of length $L$ then the thermal ensemble is the partition function of the theory on a rectangular torus with modular parameter $\tau = i \beta/L$. Furthermore, if one demands that the CFT is well defined on a torus then its partition function must be invariant under modular transformations
\begin{equation}
\tau \mapsto \frac{a \tau + b}{c\tau + d}\,, \qquad \begin{pmatrix}
a & b \\ c&d
\end{pmatrix} \in PSL(2,\mathbb{Z}) \,.
\end{equation}  

\noindent 
One of its generators is the so-called $S$-transformation $\tau \mapsto -1/\tau$. On the level of temperature, this means that the partition function must be invariant under $\beta \mapsto \frac{ L^2}{\beta}$ which directly relates different temperature regimes, in particular the partition function at very low temperature, where it is basically given by the vacuum result, to the very high temperature partition function, where high energy states dominate. This allows to bootstrap the theory and compute universal results on the growth of its high energy spectrum \cite{Nahm,Cardy:1986ie}. One can also bootstrap the CFT by considering correlation functions on the torus, which are not invariant but transform covariantly under modular transformations. This allows to obtain universal information about asymptotic matrix elements/OPE coefficients in 2d CFT \cite{km,Das:2017vej,Brehm:2018ipf}. 

A natural question while considering GGEs is whether or not similar features arise for the KdV charges. Namely, \textit{is there a transformation $y_i \mapsto \tilde{y}_i$ which relates the different regimes of the GGE and correlation functions in the corresponding state?} If such a transformation exists, can it be used to bootstrap the CFT? A related second order question is whether there is a geometrical description for GGEs such that $y_i$ plays the role of a moduli. A motivation for this expectation comes from considering classical integrable systems. Riemann surfaces play an ubiquitous role in the analytic study of integrable systems. Solutions of the integrable hierarchy are related to finding sections of an eigenvector bundle over a Riemann surface \cite{2018arXiv180506405E}. The hope is that this classical connection also reflects itself in the GGE of the quantum theory. 

Considerations of the GGE as a possible generalization of thermalization of pure states and in particular the Eigenstate Thermalization Hypothesis (ETH) also provides yet another strong motivation. It is widely believed that in many non-integrable quantum theories generic pure states look approximately thermal after waiting long enough. This, in particular, implies that static states, i.e. energy eigenstates, can be approximated by thermal states. The question arising here is whether generic pure states in integrable systems might be approximated by GGE states, rather than thermal states, when waiting long enough. This then implies that generic energy eigenstates would always look like states in equilibrium with respect to all the conserved charges of the system. 

In this article we take small steps into the direction of addressing the above questions. In particular, we compute the GGE characters in the large central charge limit. Recently in \cite{Maloney:2018yrz,Dymarsky:2018lhf,Maloney:2018hdg}, aspects of GGE was studied in the regime where the chemical potentials are parametrically smaller than the inverse temperature, which itself is small enough to probe energies much larger than the central charge. Our analysis is in a complimentary regime, which means $\mu_l \gg 1/(c^{1-\frac{1}{2l}})$, $\beta\gg 1/\sqrt{c}$, and we normally take $\mu_l\gg \beta$. Large $c$ is also the \textit{classical} limit of the corresponding soliton theory, and the integrable charges can be computed by a relatively easy recursive formula. In this limit it is possible to diagonalise all charges simultaneously and give their eigenvalues in a closed form. This allows us to give good estimates on the large and small chemical potential regimes, such that possible transformations are tractable.  

The article is organized as follows. In \S\ref{sec:SG} we review the connection between quantum Sinh-Gordon theory and conformal field theory. Following \cite{sy} we explicitly demonstrate the analytical continuation to central charges $c>1$. Our analysis reveals that the Feign-Fuchs form of the energy momentum tensor, which is needed to draw the connection to quantum integrable theory, is closely related to the bound state sectors of quantum Liouville theory that describes two-dimensional non-perturbative gravity. We will hopefully deepen the understanding of this connection in future work. In \S\ref{sec:KdVcharacters} we compute the GGE characters in mixed states with a single non-zero chemical potential at parametrically large central charge. We first concentrate on the second integrable charge and then generalize to higher ones. We again want to point out that our calculation is valid in regimes, where the chemical potentials are such that the eigenvalues that contribute dominantly to the GGE character are small compared to the central charge. In the following  \S\ref{sec:trafo} we comment on possible transformations that relate regimes of high and low chemical potentials. For this we analyze how the charge degeneracy grows within a module and perform the sum over the vacuum module for large charges by a integral evaluated via the saddle point approximation. We also compare the result to the full GGE at parametrically small chemical potentials as considered in \cite{Maloney:2018yrz,Dymarsky:2018lhf, dymarsky2}. In both cases it shows qualitatively the same behaviour. In particular this also matches the well known results of thermal states and might indicate that there indeed is some transformation that relates high and low chemical potentials. However, the results show that transformation of the chemical potentials of higher integrable charges cannot simply be the standard modular transformations and it is not clear to us if there might exist a geometric interpretation. In the following \S \ref{sec:ETH} we shortly comment on the possible generalization of ETH in the presence of integrable charges. In  \S\ref{sec:conclusion} we conclude on our findings. In the appendices we give specific formul\ae~ to compute the classical and quantum KdV charges, give some explicit checks on the large $c$ eigenvalues of the second KdV charge, and compute the global character of the second charge which appears as a factor of the GGE characters in a primary module.

\section{From CFT to quantum Sinh-Gordon theory}
\label{sec:SG}
The connection between conformal and integrable field theories has a long history. We want to go back to some of the earliest works and want to briefly review how one can connect two-dimensional conformal field theories with models like Sinh-Gordon theory \cite{sy}. Although the integrable charges themselves can be constructed for any 2d CFT the latter connection is not always present. However, only with this connection it becomes clear why large $c$ indeed is a \textit{classical} limit and how the quantum integrable structure is related to the classical integrable structure. This connection is what gives the idea to search for a geometric meaning of the chemical potentials. In what follows we do, however, not concentrate on the $c<1$ case as in \cite{sy}, but do a analytic continuation to $c>1$. In fact it turns out that the theories we have to consider have $c\ge 25$. 

Let us consider some conformal field theory and assume that we can write its energy momentum tensor in Feign-Fuchs (or Coulomb-Gas) form 
\begin{equation}\label{eq:FeiginFuchs}
T^+ (z) \equiv \frac{1}{\hbar} \left(\frac{1}{2}\no{(D\phi)^2} - \lambda D^2 \phi+ \frac{\lambda^2}2\right)\,,
\end{equation} 
where $\lambda$ is some real constant, $D = iz \partial_z$ with $z$ the compactified coordinate on the plane, $z = e^{\tau + i\sigma}$, and
\begin{equation}
\phi(z) = iq - i\alpha_0 \log(z) + i \sum\limits_{n\neq0} \frac{\alpha_n}{n}z^{-n}
\end{equation}
is a generalized free field whose modes obey the standard commutation relations
\begin{equation}
[\alpha_n,\alpha_m] = \hbar n \delta_{n+m}\,,\quad [q,\alpha_n] = i \hbar \delta_n\,.
\end{equation}

\noindent 
One can show that this energy momentum tensor corresponds to a Virasoro algebra with central charge $c = 1+\frac{12\lambda^2}{\hbar}$. Its  generators can be written as
\begin{equation}
 \hbar L_n^+ = \frac12 \sum\limits_p \no{\alpha_p\alpha_{n-p}} + i \lambda n \alpha_n + \frac{\lambda^2}{2}\delta_n\,.
\end{equation}

\noindent 
We want to follow the logic of \cite{sy} and define a second energy momentum tensor as
\begin{equation}
T^- (z) \equiv \frac{1}{\hbar} \left(\frac{1}{2}\no{(D\phi)^2} + \lambda D^2 \phi+ \frac{\lambda^2}2\right)\,.
\end{equation} 

\noindent
Hermitian conjugation is given by $(L^+_n)^\dagger = L^-_{-n}$, for which one consistent choice that we can make is the relation $\alpha_n^\dagger = -\alpha_{-n}$. Note, that by this we loose the usual direct connection between asymptotic in and out states of the quasi free field $\phi$.  

We now shortly want to comment on the similarities of the present construction to Liouville theory. Its energy momentum tensor in a strip can actually be written exactly as in \eqref{eq:FeiginFuchs} where the generalized field $\phi$ can be expressed in terms of the Liouville field $\varphi$ (see e.g. \cite{Dorn:2008sw}). When considering bound states in the Liouville theory a simple connection between in and out states by hermitian conjugation, as in the case of scattering states, is now no longer possible \cite{Dorn:2008sw}. It seems reasonable that the integrable structure that we want to present here, is also visible in the transition amplitudes of bound states in Liouville theory on a strip. 

Let us now consider the dressed Vertex operators 
\begin{equation}
V^{\pm} = z^{-\beta \lambda} \no{e^{\pm\beta\phi}}\,,\quad \beta \in \mathbb{R}\,,
\end{equation}
with conformal weight $h = - \frac{\hbar}{2}\beta^2 + \beta\lambda$ w.r.t. $T^\pm$ respectively. 
Now, to draw the connection between the CFT and the quantum Sinh-Gordon one needs $h=1$ as we will see later. The secondary field $z^{h}  V^\pm$ then naturally converges to $e^{\pm \beta\phi}$ in the classical limit $\hbar \to 0,$. Therefore, the operator
\begin{equation}
 \hat{H} = \frac{1}{2\pi i\beta^2} \oint \frac{dz}{z}\left(1-\frac{1}{2}\left(z^{1-\beta\lambda}\no{e^{\beta\phi}}+z^{1-\beta\lambda}\no{e^{-\beta\phi}}\right)\right)\,
\end{equation}
converges for $\hbar\to0$ to the classical Hamiltonian of Sinh-Gordon theory
\begin{equation}
H = \frac{1}{2\pi \beta^2}\int_0^{2\pi} d\sigma \,\left( 1-\cosh\beta\phi\right)\,,
\end{equation}
which, with the identification $u = \frac{\beta}{2}\partial_\sigma \phi$, canonically generates the Sinh-Gordon equation\footnote{The canonical Poisson bracket needs to be $\{\phi(\sigma),u(\sigma')\} = -\beta \pi \delta(\sigma-\sigma')$.} 
\begin{equation}
\partial_t\partial_\sigma \phi = \frac{2}{\beta}\partial_t u = \{u,H\} = - \frac{1}{\beta} \sinh\beta\phi\,.
\end{equation}

\noindent
An important point now is that there exist an infinite set of polynomial conserved quantities $I_n$, defined in \eqref{eq:mKdV}, which is the same as for the modified Korteweg-de-Vries (mKdV) equation 
\begin{equation}
 u_t = u_{\sigma\sigma\sigma} - 6u^2u_\sigma\,.
\end{equation} 

\noindent
Being a conserved quantity in particular means that $\{I_n,H\} = 0$. 

Another observation is that the analytic continuation of the classical Miura transform
\begin{equation}\label{eq:Miura}
v = u^2 \pm u_\sigma 
\end{equation}
maps any solution of the above mKdV equation to a solution of the KdV equation
\begin{equation}
v_t = v_{\sigma\sigma\sigma} - 6vv_\sigma\,, \label{kdv-class}
\end{equation}
which also gives rise to an infinite set of polynomial conserved quantities $K_n$ as defined in \eqref{eq:KdV}. They are connected to the conserved quantities of the mKdV by
\begin{equation}
K_n(v = u^2 \pm  u_\sigma) = I_{n+1}(u)\,.
\end{equation}

\noindent 
It is clear that 
\begin{equation}
\{K_n(v),H\} = 0\,
\end{equation}
must be true too. 
%

Let us now come back to quantum theory and the CFT. With the identification $u = \frac\beta2D\phi$ as in the classical case, the energy momentum tensor of the CFT can be regarded as a quantum Miura transform of that field,
\begin{equation}\label{eq:qMiura}
\mathcal{L}^+(z) = \hbar'\left(T^+(z) +\frac{1-c}{24}\right) = \frac{1}{2}  \left(\no{u^2} - \lambda\beta Du \right)\,,\quad \hbar' = \left(\frac\beta2\right)^2\hbar\,.
\end{equation}

\noindent
To see the mKdV hirarchy at quantum level we therefore demand that any hermitian operator/functional $\hat F(\mathcal{L}^+)$ ($\hat F(T^+)$) commutes with the quantum Sinh-Gordon hamiltonian $[\hat F(\mathcal{L}^+),\hat H] = 0$. We consider functionals of the form 
\begin{equation}\label{eq:functional}
F(T^+) = \frac{1}{2\pi i}\oint \frac{dz}{z} P^{(n)}(\mathcal{L}^+,D\mathcal{L}^+,\dots)
\end{equation}
where  $P^{(n)}$ are polynomials of $\mathcal{L}^+$ and its derivatives. All these functionals in particular commute with $V^+$ if all the Virasoro generators commute with it,
\begin{equation}
[L_n^+,z^h V^+] = 0 \quad \forall_n\,,
\end{equation}
which is only possible if the previously defined dressed Vertex operator has unit conformal weight,
\begin{equation}
 h=  - \frac{\hbar}{2}\beta^2 + \beta\lambda = 1\quad \Rightarrow \quad \lambda = \frac{1}{\beta} \left(1+\frac{\hbar}{2}\beta^2\right)\,.
\end{equation} 

\noindent
The additional constraint
\begin{align}
[F(\mathcal{L}^+),z^hV^-] = [F(\mathcal{L}^+)^\dagger,z^hV^-] =  [F(\mathcal{L}^-),z^hV^-] = 0
\end{align}
is respectively solved if for all $n$ we demand $[L_n^-,z^h V^-] = 0$ which also gives $h = 1$.

With these constraints we, in particular, obtain $c = 13+ \frac{3}{\hbar'} + 12\hbar'$, which has a minimal value of $c=25$ at $\hbar' = \frac{1}{2}$ and in the classical limit $\hbar'\to0$ goes as $c \approx \frac{3}{\hbar'}$. It also becomes evident that in the classical limit the quantum Miura transformation \eqref{eq:qMiura} becomes the classical Miura transformation \eqref{eq:Miura}. Note that the precise shift of the energy momentum tensor in \eqref{eq:qMiura} is in fact needed to preserve the same kind of connection between the KdV and the mKdV structure on quantum level as in the classical case. 
 
Following the logic in constructing the classical charges, one constructs the quantum KdV charges $\hat{K}_n$ as functionals \eqref{eq:functional}, i.e. as integrals of polynomials of $\mathcal{L}^+(z)$ and its derivatives, that are related to the mKdV charges of the quantum Sinh-Gordon theory by the quantum Miura transform \eqref{eq:qMiura}, and are constrained as follows:

\begin{itemize}
\item The charges are hermitian, $\hat{K}_n = \hat{K}_n^\dagger$.
\item The charges commute among each other, $[\hat K_n,\hat K_m] = 0$.
\item If we assign grade 2 to the field $\mathcal{L}$ and grade 1 to the derivative then $P^{(n)}$ have total grade $2n$. 
\item In the classical limit $\hbar' \to 0$ ($c\to \frac{3}{\hbar'}$) the charges  reproduce the classical integrals of motion \eqref{eq:KdV} (when identifying $v= -2\mathcal{L}$). 
\end{itemize}

\noindent
There is no closed formula known for all charges. However, one can construct them step by step with increasing grade. The first two KdV charges in a CFT are given by 
\begin{align}
  H \equiv \hat K_1 &= \frac{1}{2\pi\hbar' i}\oint \frac{dz}{z} \mathcal{L}(z) = L_0 - \frac{c-1}{24}\\
  Q \equiv\hat K_2 &=\,  \frac{1}{2\pi\hbar' i}\oint \frac{dz}{z}\, 2 \mathcal{L}^2(z) =2\hbar' \left(2\sum\limits_{n=1}^\infty L_{-n}L_n+ L_0^2 -\frac{c-1}{12} L_0 + \frac{(c-1)^2}{576} \right) \nn \\
  &=\, Q^{(1)} + 2\hbar' \left( L_0^2 - \frac{c-1}{12} L_0 + \frac{(c-1)^2}{576} \right). \label{eq:Qcharge}
\end{align}

\noindent 
More are given in \eqref{eq:CFTKdVs}\,. Note, that the charges we give here slighlty differ from the the charges considered in \cite{blz}, and also as they were used in \cite{Maloney:2018yrz,Dymarsky:2018lhf}. This is because we strictly follow the conventions of \cite{sy} and construct the charges from $\mathcal{L} = T +\frac{1-c}{24}$ to obtain the direct connection to the quantised Sinh-Gordon theory. At finite $c$ this has the effect that the charges we use, and are constructed in \cite{sy}, are some simple linear combinations of the charges used in \cite{blz}. We also want to mention that there also exists a discretized construction for the charges as given in \cite{rossi1}.

\section{KdV characters}
\label{sec:KdVcharacters}
As a first step to understand the full GGE, we start investigating the GGE characters which we define as the trace of a GGE state over a specific highest weight module,
\begin{equation}
Z^p_{\{q,y_i\}} = \Tr_{[p]}\rho(\{q,y_i\})\,.
\end{equation}

\noindent
We will analyse this quantity in the \textit{classical} limit, i.e. when $c \to \frac{3}{\hbar'} \to \infty$. We also assume that the chemical potentials are not too small which in our case means that $\mu_i \gg 1/c^{1-\frac{1}{2i}}$.\footnote{This constraint comes from a saddle point analysis and will become clear in \eqref{eq:muRange}.} This probes regimes of the GGE characters for which the corresponding charge is small compared to $c$. In this limit we are able to diagonalise all KdV charges simultaneously in a simple basis. To do so let us consider the renormalized Virasoro generators $L_n \rightarrow L_n'= \frac{L_n}{\sqrt{c}}$, with 
\begin{equation}\label{eq:Vir-norm}
[L_m',L'_n] = \frac{m^3-m}{12}\delta_{m,-n}+\frac{m-n}{\sqrt{c}} L_{m+n}' \approx \frac{m^3-m}{12}\delta_{m,-n}\,,
\end{equation}
where approximations from now on only keep the leading order results at large $c$.\footnote{This unusual normalization of the Virasoro generators with $\sqrt{c}$ also appears in \cite{Fitzpatrick:2015foa} where the authors studied the CFT dual of graviton exchanges in the heavy-light conformal blocks.} Now consider states in a primary $p$-module of the form
\begin{equation}\label{eq:states}
\ket{\{n_i,k_i\}} =\prod_{i=1}^M {L'}_{-n_i}^{k_i } \ket{p}\,.  
\end{equation}

\noindent
From the above commutation relation it follows that they are orthogonal up to large $c$ corrections. A possible basis is for example the set of all states with $L'_{-n_i}$ ordered such that bigger $n_i$ always appear on the left of smaller $n_i$. 

All the above states are in fact approximate eigenstates of all KdV charges, in the large $c$ limit. Let us first show this for the second KdV charges $Q$ as given in \eqref{eq:Qcharge}. 

\subsection{Warm-up: Q-character at large central charge}

The above states \eqref{eq:states} are clearly eigenstates of $L_0$ with eigenvalue $\sum_{i=1}^M k_i n_i$. One can show that with the commutation relation \eqref{eq:Vir-norm} it follows that they are also eigenstates of $Q^{(1)}$. This is because only terms from the central extension contribute to order $c^0$. The approximate $Q^{(1)}$ eigenvalue of a state $\ket{\{n_i,k_i\}}$ is given by
\begin{align}
 \lambda_{\{n_i,k_i\}} \approx   \sum_{i=1}^M k_i (n_i^3-n_i) \,. \label{eq:lambda}
\end{align}

\noindent
In appendix \ref{b3} we provide checks of the above equation by explicitly diagonalizing the $Q^{(1)}$ matrix for different levels in the large $c$ limit. Note that since the central term vanishes for $L'_{-1}$, contributions from possible $L_{-1}^{'m}$ to the eigenvalue are of order $1/c$. They originate from the $sl_2$ (global) part of $Q^{(1)}$, and are given by
\begin{equation}\label{eq:sl2EV}
 4 \hbar' L_{-1} L_1 {L'}_{-1}^{m } \ket{\psi} = 12 L_{-1}' L_1' {L'}_{-1}^{m } \ket{\psi} \approx \frac{12}{c} (m^2 +(2h_\psi-1)m){L'}_{-1}^{m } \ket{\psi} \,,
\end{equation}
where $\ket{\psi}$ is some state of the form \eqref{eq:states} without any $L_{-1}$, and $h_\psi$ is its $L_0$ eigenvalue. 

Note, that all possible off-diagonal elements in a basis built from \eqref{eq:states} are $o(1/c)$. 

\subsubsection{Vacuum Q character}

\textbf{First let us only consider $Q^{(1)}$:}\quad A suitable basis in the vacuum module $[0]$ is the set of all ordered states \eqref{eq:states} without any $L_{-1}'$ appearing. The vacuum $Q^{(1)}$ character is then given by
\begin{align}
\Tr_{[0]} y_1^{Q^{(1)}} &\approx \sum_{\{n_i,k_i\} \in [\Omega]} y_1^{\lambda_{\{n_i,k_i\}}} \nonumber \\
 &= \prod\limits_{n=2}^\infty \sum_{k=0}^\infty y_1^{k (n^3-n)}= \prod\limits_{n=2}^\infty \frac{1}{1-y_1^{n^3-n}}  \nonumber \\
				&= \prod\limits_{n=1}^\infty \frac{1}{1-y_1^{ 6\frac{n(n+1)(n+2)}{6} }} \label{eq:Tn1} \\
				&\equiv 1+\sum_{m=1} P_{\{T_n\}}\!(m) \,x^m \,,\quad\text{with } x = y_1^{6}.\label{eq:Tn2}
\end{align}

\noindent
The numbers $T_n = \frac{n(n+1)(n+2)}{6}$ are the tetrahedral numbers. They are defined as the number of balls needed to construct a tetrahedral cluster with the close-packing of equal spheres. From \eqref{eq:Tn1} one can hence show that the vacuum $Q^{(1)}$-character is a generating function for the number of partitions of an integer $n$ into tetrahedral numbers $P_{\{T_n\}}(m)$. We will see similar partitions for higher charges. 

\medskip\noindent
\textbf{Other relevant term in $Q$:}\quad The other non-constant term relevant at large $c$ is $-\frac{\hbar'(c-1)}{6} L_0 \approx -\frac{L_0}{2}$. Taking it into account one obtains 
\begin{equation}
\Tr_{[0]} y_1^{Q} \approx \prod\limits_{n=2}^\infty \sum_{k=0}^\infty y_1^{k (n^3-\frac32n)}= \prod\limits_{n=2}^\infty \frac{1}{1-y_1^{n^3-\frac32n}} 
\end{equation} 

\noindent
The $Q^{(1)}$ eigenvalues get shifted by the (scaled) energy of the corresponding states and at least for the low lying eigenvalues a lot of the tetrahedral degeneracy is broken. However for large $n$, the partitions still grow exponentially in the same way as the partition into tetrahedral numbers.  

\subsubsection{Primary Q character}

\textbf{Again first let us consider $Q^{(1)}$:}\quad As mentioned earlier the possible $L_{-1}'$ in the states contribute sub-leading to the eigenvalues with \eqref{eq:sl2EV}. For states in a primary module only containing $L_{-1}'$ the $Q^{(1)}$ eigenvalue is even sub-dominant. Taking this into account we can write the $Q^{(1)}$-character in the primary $p$ module, $[p]$, as
 \begin{align}
\Tr_{[p]} y_1^{Q^{(1)}} &\approx \prod\limits_{n=2}^\infty \sum_{k=0}^\infty y_1^{k (n^3-n)}+\left(\sum_{k=1}^\infty y_1^{\frac{12}c \left(k(2p-1) +k^2\right)}\right) \prod\limits_{n=2}^\infty \sum_{k=0}^\infty y_1^{k (n^3-(1-\frac{24}{c})n)} \nonumber \\
&\approx \left(\sum_{k=0}^\infty y_1^{\frac{12}c \left(k(2p-1) +k^2\right)}\right)\prod\limits_{n=2}^\infty \frac{1}{1-y_1^{ (n^3-n)}}
\end{align}

\noindent 
This result is the vacuum result \eqref{eq:Tn1} multiplied with the global character result, as given in \eqref{eq:globalQ1}. 

\medskip\noindent
\textbf{Full $Q$:}\quad Again we can write the result as the vacuum character multiplied with the global character as given in \eqref{eq:FullQsl2}. It is given by
\begin{align}
\Tr_\text{p} y_1^{Q} &\approx \left(\sum_{k=0}^\infty y_1^{\frac{6(p+k)^2 + 12 (k^2 +2pk-k)}{c}-\frac{1}{2} (p+k)}\right)\prod\limits_{n=2}^\infty \sum_{k=0}^\infty y_1^{\frac{1}{2}k (n^3-\frac32n)} \nonumber \\
&= \left(\sum_{k=0}^\infty y_1^{\frac{6(p+k)^2 + 12 (k^2 +2pk-k)}{c}-\frac{p+k}{2} }\right) \prod\limits_{n=2}^\infty \frac{1}{1-y_1^{\frac{1}{2} (n^3-\frac32n)}} \label{primary-sum}\,.
\end{align}

\noindent
One may estimate the term in the parentheses by extremizing the exponent with respect to $k$ and approximate the sum by the corresponding $k$ saddle value, see \S \ref{Q-closed}. Such an analysis however, is expected to be quite different from the case when $c$ is fixed, since in that case the dropped terms $\propto 1/c$ will contribute significantly to the character.

%

\subsection{Higher charge characters}
\label{sec:higherK}
We now want to analyze all the higher charges in the large $c$ limit. As mentioned before, this is the classical limit which in particular means that all the quantum corrections to the classical KdV charges/currents can be neglected. While a closed form doesn't exist for the quantum charges,  the classical charges are given explicitly in \eqref{eq:KdV}. To compute the quantum charges in the large $c$ limit we simply have to replace $v$ in \eqref{eq:KdV} by $-2\mathcal{L}$, multiply with $(-1)^k$ to obtain a positive expression, and consider the respective symmetric composite field. However, even that can be reduced further. In the classical limit we can write $\mathcal{L} = \hbar' T - \frac{1}{8}$ (since $c \to 3/\hbar'$ for $\hbar'\to 0$) and the $k$th current ($\equiv \mathcal{I}_k$) can be written in a power series in $\hbar'$, 
\begin{equation}
(-1)^k \hbar'\mathcal{I}_k \approx \sum_{n=0}^k P^{(n)}_k\!(T,D^2T,\dots)\hbar'^n\,,\label{eq:hexpansion}
\end{equation}
where $P^{(n)}_k\!(T,D^2T,\dots)$ is some polynomial of grade $2n$ of $T$ and its derivatives. $T$ contributes two and any derivative one to the grade. It follows that $P^{(n)}_k$ and $P^{(n+1)}_k$ can at most give contributions to the eigenvalues of order $c^{n/2}$, where the eigenvectors are still given by \eqref{eq:states}. We can therefore neglect all terms in \eqref{eq:hexpansion} with $n>2$ since they contribute at least of order $c^{-1/2}$ to $\mathcal{I}_k$. Therefore, the $k$th current ($k>1$) can be approximated by  
\begin{equation}
(-1)^k \hbar'\mathcal{I}_k \approx  \frac{Ct\!(k-1)}{2^{2k+1}} + P^{(1)}_k\!(T)\,\hbar' + P^{(2)}_k\!(T,DT,\dots)\,\hbar'^2 \,. 
\end{equation}

\noindent
$Ct\!(k)$ is the $k$th Catalan number and follows from the recursive definition of the classical charge. $P^{(1)}_k\!(T)$ has the form 
\begin{equation}
P^{(1)}_k=\frac{- \sqrt{\pi}}{(k-1)! \left|\Gamma(\frac{3}{2}-k)\right|} T(z) + ~\text{total derivatives},
\end{equation}
and therefore contributes
\begin{equation}
 \frac{-\sqrt{\pi}}{(k-1)! \left|\Gamma(\frac{3}{2}-k)\right|} L_0
\end{equation}
to the charge.

The remaining polynomial can be written as
\begin{equation}
P^{(2)}_k =  \sum_{m=0}^{k-2}  b_m^k T(z)D^{2m}T(z) \,\quad \text{for } k\ge 2\,,
\end{equation}
with $b_m^k$ positive constants, where always $b_{k-2}^k=2$. The other coefficients can e.g. be computed by the help of a computer algebra program. Integrating over this part of the current then gives
\begin{equation}
 \frac{1}{2\pi i \hbar'} \oint \frac{dz}{z} P_k^{(2)} \hbar'{}^2 = 2 \hbar' \sum_{n>0} \left(\sum_{m=0}^{k-2}(-1)^m b_m^k n^{2m}\right)  L_{-n}L_n  \,,
\end{equation}
where for now we omit the term proportional to $L_0^2$ which is sub-leading in large $c$. The relevant part of the $k$th charge in the large $c$ limit can therefore be written as
\bea
(-1)^k \hat{K}_k &=& 2 \hbar'  \sum_{n>0} \left(\sum_{m=0}^{k-2} (-1)^mb_m^k n^{2m}\right)  L_{-n}L_n -  \frac{\sqrt{\pi}}{(k-1)!\left| \Gamma(\frac{3}{2}-k)\right|} L_0 +  \frac{Ct\!(k-1)}{2^{2k+1}\hbar'}, \nn \\
&\equiv& K_k^{(1)} -  \frac{\sqrt{\pi}}{(k-1)! \left|\Gamma(\frac{3}{2}-k)\right|} L_0 + \frac{Ct\!(k-1)}{2^{2k+1}\hbar'}. \label{Kk}
\eea

\noindent
With this it is clear that the previously chosen orthogonal basis in large $c$ diagonalises all the higher charges as well. Firstly, note that we have only kept the leading order terms in large $c$. At $\mathcal{O}(1/c)$, the only contributions come from the $L_0^2$, and the $L_{-1}L_1$ term in $K_k^{(1)}$. Secondly, note that the most suppressed term is 
\be
(-1)^k \hbar^{k-1} 2^{k-1} Ct\!(k-1) L_0^k \approx (-1)^k \frac{6^{k-1}}{c^{k-1}}Ct\!(k-1)\, L_0^k. \label{eq:L0k}
\ee

\subsubsection{$\hat{K}_k$ vacuum character}
Now, we focus on the vacuum module. Using \eqref{Kk} we obtain, 
\begin{equation}
\hat{K}_k \ket{\{n_i,l_i\}} = \mathfrak{K}_{k,\{n_i,l_i\}}\ket{\{n_i,l_i\}}
\end{equation}
with
\begin{align}
\mathfrak{K}_{k,\{n_i,l_i\}} &= \sum_{i=1}^M l_i \bigg[ \left(n_i^3-n_i\right) \left(\sum\limits_{m=0}^{k-2}(-1)^{k+m}\frac{b_m^k}{2} n_i^{2m}\right) - \frac{(-1)^k\sqrt{\pi}\,n_i}{(k-1)! \left|\Gamma(\frac{3}{2}-k)\right|} \bigg] +  \frac{(-1)^k C\!t\!(k-1)}{2^{2k+1}\hbar'}\,\nonumber\\
&\equiv  \sum_{i=1}^M l_i\left[ \mathfrak{p}_{k,n_i}^{(1)}- \frac{(-1)^k\sqrt{\pi}\,n_i}{(k-1)! \left|\Gamma(\frac{3}{2}-k)\right|}\right] +  (-1)^k\frac{C\!t\!(k-1)}{2^{2k+1}\hbar'}\,. \label{eq:KdVEV}
\end{align}

Focussing on the $K^{(1)}$ part of the $l$-th KdV charge, we find the character analogous to \eqref{eq:Tn1} as, 
\begin{align}
\Tr_{[0]}  y_l^{K_l^{(1)}} &\approx \prod_{n=2}^\infty \sum_{k=0}^{\infty} y_l^{k\cdot \mathfrak{p}_{l,n}^{(1)}} \nn \\
&= \prod_{n=2}^\infty  \frac{1}{ 1 - y_l^{\mathfrak{p}_{l,n}^{(1)} }  }.
\end{align}

\noindent
For $\mu_l \ll 1$ ($\mu_l = - \log y_l$) we can assume that the above is dominated by large $n\sim n_* \gg1$.\footnote{ However, since in our approximation $c \gg n_*$, we will also obtain a lower bound on $\mu_l$.} For large enough $n$ we can approximate $\mathfrak{p}_{l,n_i}^{(1)} \approx n^{2l-1}$ and , hence, obtain
\begin{equation}
\Tr_{[0]}  y_l^{K_l^{(1)}} \approx 
\prod_{n=2}^\infty \left( 1 - y_l^{n^{2l-1} }\right)^{-1}\,. \label{intchar}
\end{equation}
Remarkably, \eqref{intchar} can be expressed in an expansion in $y_l$ weighted by power partitions, $p_l(q_m)$, 
\begin{equation}
 \Tr_{[0]}  y_l^{K_l^{(1)}}\approx  \sum_{q_m=0 }^\infty p_l(q_m) y_l^{q_m}\,. \label{charl}
\end{equation}

\noindent 
$p_l(q_m)$ is the number of ways $q_m$ can be expressed as sum of integers which are of the form $i^{2 l -1}$ where $i$ is also an integer (this is the quantum number of the $l$th KdV charge). Note that for $l=2$, this is the partition of $n$ into cubes, which is the large $n$ asymptotics of tetrahedral numbers. It turns out that there is a formula due to Hardy and Ramanujan \cite[p.111]{Hardy1918} which gives an asymptotic analytic form for the power partitions,
\be
p_l(q_m) \approx \exp \left( A(l)\, q_m^{\frac{1}{2l} } \right), \label{popa}
\ee 
where, 
$
A(l)= 2l \left( \frac{1}{2l-1} \Gamma( \frac{2l}{ 2l-1})\zeta( \frac{2l}{ 2l-1}) \right)^{(2l-1)/(2l) }$. This then gives us the density of states in the vacuum module with $K^{(1)}_l$ charge $q_m$, where $ q_m \gg 1$. Adding the $L_0$ contribution does not change the above logic which allows us to approximate the vacuum character by 
\begin{equation}
 \Tr_{[0]}  y_l^{K_l} \approx \sum_{q_m=0}^\infty \exp\left(A(l)q_m^{\frac{1}{2l}} - \mu_l \left(q_m -  c B_l\right)\right)\,,
\end{equation}
where $B_l = (-1)^{l+1}\frac{C\!t\!(l-1)}{3\,2^{2l+1}}$ comes from the constant part in the charge. We can further approximate the sum by an integral and obtain 
\begin{align}
\Tr_{[0]}  y_l^{K_l} &\approx e^{\mu_l c B_l} \int_0^\infty dq_n \exp\left(A(l)q_n^{\frac{1}{2l}} -\mu_l q_n \right)\\
&\approx e^{\mu_l c B_l}\exp\left\{  \frac{2l-1}{2l} A(l)^{\frac{2l}{2l-1}} \left(2l \mu_l\right)^{\frac{1}{(1-2l)} } \right\}\,, \label{highMu}
\end{align}
where the integral has been done by saddle in \eqref{saddle1}.

\subsubsection{$\hat{K}_k$ primary characters}

\textbf{Only $K^{(1)}_l$:\quad} As in the case of $Q^{(1)}$ we only have to consider the additional contributions from $L_{-1}L_1$ in the sum. It is sub-leading and adds 
\begin{equation*}
\frac{24}c  \left(k(2(p+N)-1) +k^2\right) \mathfrak{A}
\end{equation*}
to the dominant contribution, where $N$ denotes the contribution to the conformal dimension from all excitations not including $L_{-1}$ and $\mathfrak{A} = \sum_{m=0}^{l-2}(-1)^m b_m^l$. Analogously to $Q^{(1)}$ we can therefore write
\begin{align}
\Tr_{[p]}  y_l^{K_l^{(1)}} &\approx \prod_{n=2}^\infty \sum_{k=0}^{\infty} y_l^{k \mathfrak{p}_{l,n}^{(1)}} +  \left(\sum_{k=1}^\infty y_1^{\frac{12}c  \left(k(2p-1) +k^2\right)\mathfrak{A}}\right)\prod_{n=2}^\infty \sum_{k=0}^{\infty} y_l^{k \left( \mathfrak{p}_{l,n}^{(1)}+\frac{24}c \mathfrak{A} n\right)} \nn \\
&\approx \left(\sum_{k=0}^\infty y_1^{\frac{12}c \mathfrak{A} \left(k(2p-1) +k^2\right)}\right) \prod_{n=2}^\infty  \frac{1}{ 1 - y_l^{\mathfrak{p}_{l,n}^{(1)} }  }.
\end{align}

\noindent
Again, this is the global answer multiplied with the vacuum result.

\noindent
\textbf{Full $K_l$:\quad} Again, analogous to the case of $Q$, the dominant contributions from $L_{-1}$ to states in the $p$-module come from the part in $\hat{K}_l$ proportional to $L_0$. Formally, we can write 
\begin{align}
\Tr_{[p]}  y_l^{K_l} &\approx  \left(\sum_{k=0}^\infty y_1^{ \frac{-\sqrt{\pi}}{(k-1)! \left|\Gamma(\frac{3}{2}-k)\right|} (p+k)+\frac{12}c  \left(k(2p-1) +k^2\right)\mathfrak{A}+\frac{b^l_0}{c} (p+k)^2}\right)\prod_{n=2}^\infty  \frac{1}{ 1 - y_l^{\mathfrak{p}_{k,n}^{(1)}+ \frac{-\sqrt{\pi}\,n}{(k-1)! \left|\Gamma(\frac{3}{2}-k)\right|}   }  }\,.
\end{align}


\subsection{Comment on the full GGE}

The full Generalized Gibbs Ensemble is obtained by taking the sum over all GGE characters that appear in the CFT. This sum, however, depends on the specifics of the theory. Note that we cannot approximate the result by e.g. taking the asymptotic formula for the growth of primary states \cite{Nahm, Cardy:1986ie}, because this formula is valid in energy regimes ($\Delta_p\gg c$) that do not fit the regimes of validity that we consider ($\Delta_p\ll c$). We would need the specifics of the spectrum at intermediate energies which is not universal \cite{Hartman:2014oaa}. 

\subsection{Constraints on $\mu_l$}\label{mu-const}

We already mentioned it earlier and now want to show that one gets a lower bound on $\mu_l$ by demanding that the dominant contribution from the saddle point of the above integral is smaller than $c$. The saddle turns out to be,
\begin{align}
1 &\ll q_n^* = 2^{\frac{2l}{1-2l}} \left(\frac{l \mu_l}{A(l)}\right)^{\frac{2l}{1-2l}} \ll c\nonumber\\
\Rightarrow \qquad \mu_l &\gg \frac{A(l)}{2l} \frac{1}{c^{1-\frac{1}{2l}}} \approx \frac{1}{c^{1-\frac{1}{2l}}}\,. \label{eq:muRange}
\end{align}
This implies that our result \eqref{highMu} is valid in the regime
$$
c^{\frac{1}{2l}-1} \ll \mu_l \lll 1\,.
$$

\section{Connection between high and low chemical potentials}
\label{sec:trafo}
As mentioned in our introduction one of our motivations to study integrable charges and the GGE is to reveal some possible transformation property of the chemical potential under which the full GGE or GGE characters behaves in a certain way. In this section we show that our results allow to discuss some aspects of the transformation. 

First, let us collect the results at large and small chemical potential and large $c$ with a single integrable charge $\hat{K}_l$. At high chemical potential the corresponding GGE is dominated by the constant part of the charge because it has the highest power in $c$. However, the contribution is not necessarily given by the vacuum contribution because the charges are not necessarily positive. The state with lowest charge eigenvalue dominates. Nevertheless we can only restrict ourselves to the constant part, because all other contributions in this regime are hugely sub-dominant. So at large chemical potential $1\ll\mu_l\ll c$, and with all other chemical potentials turned of, we can approximate the GGE by 
\begin{equation}
Z(\mu_l) \approx e^{\mu_l c B_l}\,,
\end{equation}
where $B_l = (-1)^{l+1} \frac{Ct\!(l-1)}{2^{2l+1}} $ is the constant part computed in the last section. 

In the regime of small chemical potential, $1/c^{1-1/2l}\ll \mu_l \ll 1$, we in particular computed the $\hat{K}_l$ vacuum character in \eqref{highMu}, 
\begin{align}
\Tr_{[0]} y_l^{\hat{K}_l} &\approx e^{\mu_l c B_l}\exp\left\{  \frac{\mathcal{C}_l}{ \mu_l^{\frac{1}{2l-1} }} \right\}\,,
\end{align}

\noindent 
where $\mathcal{C}_l = \Gamma \bigg( \frac{2l}{2l-1} \bigg) \zeta \bigg( \frac{2l}{2l-1} \bigg)$. The factor in the front, $e^{\mu_l c B_l}$, again comes from the constant part. However in the case of small chemical potentials we can show that this factor is negligible for large $l$. The competition is between the factors, $\mu_l c B_l$ and $\mathcal{C}_l \mu_l^{\frac{1}{1-2l}}$. When $\mu_l$ is taken the smallest it can be (to be in the valid regime of our analysis, i.e, $\mu_l \sim c^{-1+\frac{1}{2l}}$), the two terms are of the order, $\sqrt{c^l} B_l$ and $\sqrt{c^l}\mathcal{C}_l$. Therefore we need to compare $\mathcal{C}_l$ and $B_l$. For large $l$, using the asymptotics of Gamma and Zeta functions, $\mathcal{C}_l \sim  2l - 1$ whereas, using the asymptotics of Catalan numbers, $|B_l| \sim l^{-3/2}/( 8\sqrt{\pi})$. Therefore clearly, for small $\mu_l$, as $l$ gets bigger, 
\begin{align}
\Tr_{[0]} y_l^{\hat{K}_l} &\approx \exp\left\{  \frac{\mathcal{C}_l}{ \mu_l^{\frac{1}{2l-1} }} \right\}. \label{smallchem}
\end{align}


If we now assume that there exist any nice transformation property that relates the low and high chemical potentials relating the respective GGEs, it is reflected in the second factor $\exp\left\{  \mathcal{C}_l/ \mu_l^{\frac{1}{(2l-1)} } \right\}$. Since the vacuum character appears in all primary character we can assume that this factor multiplies the full GGE. 

It follows immediately that, if $\mu_i\neq \beta$, it is not possible to connect the low and high potential regime by an '$S$ transformation', $\mu_i \rightsquigarrow \frac{1}{\mu_i}$. Rather, the transformation that connects the high and low chemical potential regime seems to have the form
\begin{equation} \label{eq:mutrafo}
\mu_i    \rightarrow  \mathcal{D}_i \mu_i^{\frac{1}{1-2i}}\,,
\end{equation}
where $\mathcal{D}_i$ is independent of $\mu_i$, just depending on the charge and $c$. 
\noindent
It is clear that the GGE can definitely not be invariant under this transformation. This is because any function of $\mu_i$ that is invariant under such a transformation would be constant. This follows easily by performing $\mu_i \rightarrow  \mathcal{D}_i \mu_i^{\frac{1}{1-2i}}$ several times. 
\begin{equation}
f(\mu_i) = f\left(\mathcal{D}_i^{\sum_{j=0}^{k-1}\frac{1}{(1-2i)^j} }\mu_i^{\frac{1}{(1-2i)^k}}\right)\qquad \forall k\in \mathbb{N}, \mu_i \in \mathbb{R}^+\,.
\end{equation}

\noindent 
Taking the limit $k\to\infty$ then tells that $f(\mu_i) = f(\mathcal{D}_i^{\frac{2i-1}{2i}})$.

\subsection{Connection to very high energies}
Recent study of KdV in 2D CFTs have explored various aspects of the GGE in the regimes of asymptotically large energies and KdV charges  \cite{Maloney:2018yrz,Dymarsky:2018lhf, Maloney:2018hdg}. In particular, the chemical potentials are outside the regime that we have studied, viz., $\mu_i \ll \frac{1}{c^{1-1/2i}}$. Therefore in this regime, the GGE partition function is dominated by $h_*, n_* \gg c$. Additionally, the investigations also focus on the limit, $\{\mu_l \} \ll \beta \ll 1$, therefore the KdV charges are perturbatively treated on the torus. It was shown that when high energies dominate, the main contribution to the GGE comes from decendents at particular level of primaries at a particular energy regime, and the GGE can be approximated by (using \eqref{eq:L0k})\footnote{ Our coefficients in the exponent differ from those in \cite{Maloney:2018yrz,Dymarsky:2018lhf, Maloney:2018hdg}, since the expressions for KdV charges (for $c>1$) are different. This is because we have chosen the shift to $T(z)$ to be $\tfrac{1-c}{24}$ as in \cite{sy} while in \cite{blz,blz2,blz3} the shift is $-\tfrac{c}{24}$, which is followed by \cite{Maloney:2018yrz,Dymarsky:2018lhf, Maloney:2018hdg}. However, the large $c$ expressions of course match! },
\begin{equation}
Z(\{q,y_i\}) = \Tr q^H \prod_i y_i^{\hat{K}_i} \approx \int_0^\infty dh\, \rho(h) \exp\left(-\beta h - \sum_i \mu_i \frac{ 6^{i-1} C\!t(i-1)}{c^{i-1}}  h^i\right)\,.
\end{equation}

\noindent
In this high energy regime unlike \cite{Maloney:2018yrz,Dymarsky:2018lhf, Maloney:2018hdg}, we shall focus on $1 \gg \{\mu_l \} \gg \beta$. The integral can in principle be solved by a saddle point approximation. This, however, becomes very complicated for arbitrary $\mu_i \ll 1/c^{1-\frac{1}{2i}}$. The situation simplifies when we take all potentials parametrically small with respect to one particular potential. In that case, the saddle is always dominated by the contribution of that particular potential with corrections given by a power series in all the other ones. The dominant saddle $h^*$ for $\mu_n\gg \{\mu_i\}, \beta$  is given by
\begin{equation}
h^* = \left(\frac{\pi}{ n 6^{i-1/2}C\!t(i-1) \mu_n }\right)^{\frac{2}{2n-1}} c + \sum_{i\neq n} C_i \mu_i + \mathcal{O}(\mu_i \mu_j) \,,
\end{equation}
where the coefficients $C_i \equiv C_i(\mu_n,c)$ are such that the corrections are small. Plugging back this saddle, the partition function is approximately given by
\begin{equation}\label{eq:lowMuMaloney}
 Z(1 \gg \mu_n\gg \mu_i) \approx \exp\left[\frac{(2n-1)c}{6}\left(\frac{ \pi^{2n}}{C\!t(n-1) n^{2n} \mu_n}\right)^{\frac{1}{2n-1}} + \mathcal{O}(\mu_i)\right]\,.
\end{equation}

\noindent 
Thus a bit surprisingly, we see that the GGE for very low chemical potential $\mu_l\ll \frac{1}{c^{1-1/2l}}$ shows the same $\mu_l$ dependence that we saw when we analyzed the universal vacuum character in the range $\frac{1}{c^{1-1/2l}} \ll\mu_l\ll 1$, \eqref{smallchem}. 

This is also compatible with the known results from the thermal partition function. The Virasoro characters and the full partition function at high temperature show qualitatively the same behavior. The sum over all characters must lead to a quantitative agreement and, hence, gives tight constraints on the spectrum of the theory. The latter e.g. allows to compute the asymptotic growth of the density of primary states. In the GGE case, using \eqref{eq:lowMuMaloney} we can derive the growth of the asymptotic density of states carrying the $n$-th KdV charge ($q_n$) as, $\exp\left( A q_n^{\frac{1}{2n} }\right)$ (same $q_n$ dependence as in \eqref{popa}).

\section{Comment on approximations of pure states}
\label{sec:ETH}
Studying the GGE can also shed more light on how the eigenstate thermalization hypothesis (ETH) \cite{srednicki1996thermal,Deutsch:1991,rigol2008thermalization} might be more generally realized in two-dimensional conformal field theory. It states that for a a particular class of simple operators generic energy eigenstates can be well approximated by thermal states. It is, however, only expected to be true in theories exhibiting chaotic dynamics and known to fail in integrable theories, where integrable here not only means that there exists an infinite set of conserved charges but also that these trivialize the dynamics. In case of 2d CFTs with $c>1$ we saw that there exist an infinite set of conserved charges. However, the dynamics is not necessarily trivial. For example in \cite{Roberts:2014ifa} it has been shown that 2d CFTs exhibit chaotic dynamics and, hence, a generalization of ETH might hold. The generalization takes not only the Hamiltonian/energy into account but considers all conserved quantities. One might expect that a generic static state can for some class of simple operators by approximated by some equilibrium state. The latter defines a GGE by taking the trace over the spectrum of the theory. 

We, here, want to check how well a GGE dominated by a single chemical potential, $\mu_l$, can approximate a typical eigenstate of the corresponding KdV charge, $\hat{K}_l$, in the regime that we probe. For now let us choose such a state $\ket{q_m}$, whose $\hat{K}_l$ eigenvalue is $q_m- cB_l$, where $1 \ll q_m\ll c$. Restricting to the vacuum part of the GGE, 
in the regime $\{\mu_i, \beta \} \ll \mu_l\ll1$, then we can compute $\langle\hat{K}_l \rangle$ in the mixed state $e^{-\mu_l \hat{K}_l}$ reduced to the vacuum module with the help of \eqref{highMu} as
\begin{align}
\langle \hat{K}_l\rangle_{q_l} &= -\partial_{\mu_l} \log Z^{0,l}_{GGE} \nn\\
&\approx - \partial_{\mu_l} \left\{  \frac{2l-1}{2l} A(l)^{\frac{2l}{2l-1}} \left(2l \mu_l\right)^{\frac{1}{1-2l}} + \mu_l c B_l \right\}\nn\\
&=    \left(2l A(l) \mu_l \right)^{\frac{2l}{1-2l}}  -  cB_l\,.
\end{align}

\noindent
Hence, a necessary condition for $e^{-\mu_l \hat{K}_l}$ being a good approximation to $\ket{q_m}\bra{q_m}$ is when we equalize the eigenvalue, $q_m - c B_l$ to the above ensemble answer. This gives, 
\begin{equation}
 \mu_l = \frac{1}{2lA(l)} q_m^{\frac{1-2l}{2l}}\,. \label{eq:muq}
\end{equation}

\noindent
Note that since the GGE should be a positive quantity, $\mu_l$ should be a real quantity, implying positivity of $q_m$. Thus in the cases where $q_m <0$ it appears that we cannot have equilibriation. In the regime we consider, this can e.g. happen if the state happens to have zero excitations or a lot of light excitations viz. $L_{-1}^k$. Then the term proportional to $\sum_{n>1}L_{-n}L_n$ contributes not enough to the charge to overcome the negative contribution from the term proportional to $L_0$. This also is an artifact of the large $c$ regime we are working in, since higher powers of $L_0$ (which are subleading in $\mathcal{O}(1/c)$) can make $q_m >0$ when one goes beyond our regime and in particular to much higher energies. 

It is interesting to note, that states close to the primary, in the sense that they are almost no excitations on top of it, can be well approximated by a purely thermal state in the large $c$ limit \cite{Guo:2018pvi}. However, in \cite{Guo:2018pvi} they also conjecture -- and show it for particular states explicitly -- that descendants far from their primary state cannot be approximated well by a purely thermal state, even at large $c$! For such states, $q_m \gg 0$ and GGE states might be a necessity for a good approximation of those. 

\section{Conclusions}
\label{sec:conclusion}
With this note we have shed some light on the implications that follow from the integrable structure of 2D CFT. We have first shown under which conditions one can not only construct an infinite set of commuting charges in a CFT with $c>1$, but also connect it to the quantum Sinh-Gordon theory. Our analysis implies that if one can write the energy momentum tensor in the Feigin-Fuchs form, then the quantum KdV charges of CFT with $c = 13 + \frac{3}{\hbar'} + 12\hbar\ge 25$ are related to the quantum mKdV charges of sinh-Gordon (sG) theory with the Hamiltonian 
\begin{equation}
\hat{H} = \frac{1}{2\pi i} \oint \frac{dz}{z} \left(1-\frac{1}{2z^{\frac{\hbar'}2}}\left(\no{e^{\beta\phi}}+\no{e^{-\beta\phi}}\right)\right).
\end{equation}
The relation between the quantum KdV and quantum mKdV charges are given by the Miura transform \eqref{eq:qMiura}. 

Next we have concentrated on the KdV \textit{characters} at large central charge which on the sG side corresponds to taking the classical limit ($\hbar \sim 3/c \rightarrow 0$). They are the building blocks of generalized Gibbs ensembles and can help in understanding its general features. The KdV characters can be computed for any integrable charge $\hat{K}_l$, as long as the corresponding chemical potential $\mu_l$ is bounded from below by $1/c^{1-\frac{1}{2l}}$. This probes regimes where the corresponding eigenvalues of the charge are small compared to the central charge, which is always kept parametrically large in our analysis. The key idea is that to leading order in large $c$ all states of the form \eqref{eq:states} diagonalize the charges with eigenvalues given by \eqref{eq:KdVEV}. 

A key motivation behind this work, has been whether there exists a connection between the high and low chemical potential GGE. By using the Hardy-Ramanujan result \eqref{popa} on the number of partitions into monomials we can find the behaviour of the KdV vacuum character at small chemical potential \eqref{highMu}. This factor appears in the other characters as well, and is universal. The self-consistency of our analysis bounds the chemical potential $\mu_l$ from below by, $1/c^{1-\frac{1}{2l}}$. When considering much lower chemical potentials, s.t. eigenvalues much larger than $c$ dominate the GGE, one can also extract the result for the full GGE \eqref{eq:lowMuMaloney} from the results in \cite{Maloney:2018yrz}. Both methods show the same $\mu_l$ dependence in the GGE answer, which suggests that this behavior is valid for a rather large range of $\mu_l\ll1$. On the other end, for large $\mu_l$, the GGE is dominated by the vacuum. In case of the first KdV charge, namely the Hamiltonian, the two regimes are directly related by the standard $\mathcal{S}$-transformation, which takes $\beta \rightarrow 1/\beta$. It turns out, that for the higher chemical potentials a transformation of the form, $\mu_l \rightarrow \mu_l^{\frac{1}{1-2l}}$, \eqref{eq:mutrafo} relates the low and the high $\mu_l$ GGEs .

We also briefly commented on a generalization of the eigenstate thermalization hypothesis. If a thermal state fails to approximate an energy eigenstate it might be cured by considering GGE states. Once again in the restricted regime of a single chemical potential dominated GGE, we show the conditions when such an approximation is well defined. States that are already known to be approximated well by a thermal state, namely high energy primary states and descendant states with only light excitations on top of the primary \cite{Guo:2018pvi}, can - at least in the large $c$ regime of validity of our analysis  - \textit{not} be approximated by a GGE state with only a single charge other than the Hamiltonian turned on, since the corresponding chemical potential is then no longer \textit{real}. On the other hand for pure states made out of large excitations above the highest weight state, one can solve for the GGE chemical potential that needs to be turned on \eqref{eq:muq}. 

A number of interesting directions are left for future work. Here we list some of the more pressing ones,
\begin{itemize}
\item going beyond the large $c$ approximation : to proceed in the $1/c$ calculation of the GGE character, one needs to know more sub-leading terms in $K_k$ beyond the regime of the present analysis. In future work, we shall try to see if the issue of finite $c$ can be addressed by extending the technologies developed in \cite{blz,blz2,blz3} to $c>1$. Since the problem of finding the eigenvalues of all KdV charges in the CFT can be mapped to the problem of solving a particle in a specific potential it might also be possible that symmetries of the GGE are directly related to symmetries in the spectrum of the quantum particle. 
\item investigate the GGE in specific models :  until now, the GGE with only the second charge turned on, has not even been computed for free theories. We in particular hope, that a closed form of the GGE might reveal more general transformation properties of the chemical potentials and show that the connection between high and low chemical potential that we see in this paper is not an artifact of taking large $c$. For free bosons, the largest eigenvalue of $K_2$ at level $n$ seems to grow as $n^3$ \cite{samarth}. 
\item issues of equilibriation : the pure states in 2D CFT are of the form, $\ket{\psi} = \prod_{i} L_{-n_i}^{k_i} \ket{p}$ where, $\ket{p}$ is a highest weight state, with $L_0$ eigenvalue $E = p + \sum_i n_i k_i$. Hence the thermal ensemble is only sensitive to the number $E$. On the other hand, to fix a KdV GGE one has to use more information regarding the construction of the state. There are clear indications of this even in large $c$. For example, the large $c$, eigenvalue $\lambda_{\{n_i, k_i\}}$ of the second KdV charge, \eqref{eq:lambda}, is not just the combination $E$, and thus probes a finer microstructure of $\ket{\psi}$. Higher KdV charges probe even finer microstructures, however to understand this completely, one will need to go beyond large $c$, since in leading order in $1/c$, the $\lambda_{\{n_i, k_i \}}$ eigenvalues continue to dominate. 

\item holographic interpretation : adding polynomials of the stress tensor and its derivatives with respective chemical potentials to the action of a CFT corresponds to a multitrace deformation in the CFT. In a potential dual description of the CFT these deformations also appear as multitrace deformation of the action of gravitational action (or its higher spin extension). As discussed in \cite{deBoer:2016bov} these can for example be studied in the Chern-Simons formulation. The deformation appears as a boundary term which does not change the bulk field equations and hence a classical BTZ black hole solution stays unchanged. However, the free energy/partition function is different from the that of a usual BTZ black hole because of the additional boundary terms. In fact, the change looks like a generalized Legendre transform which changes the partition function to the so-called tau-function of the classical KdV hierarchy. When quantizing the Chern-Simons theory with matter this leads to more structure in the gravitational solution and could lead to explicit constructions of black holes with quantum hair. 

\end{itemize}

Apart from the above, it will also be interesting to understand the geometric interpretation of the KdV charges, in the same way that the Hamiltonian translates states along the Euclidean time. A generalization of the mapping of GGE to a quantum mechanics problem \cite{blz,blz2, blz3} for $c >1$ and for arbitrary chemical potentials is another fascinating future direction.

\section*{Acknowledgements}
It is a pleasure to thank George Jorjadze, and Stefan Theisen for encouragement and useful discussions at various stages of the project. We would also like to thank Shouvik Datta who was part of the initial collaboration. DD is grateful to the MPI Partner group grant.

\section*{Attachments}

We uploaded two Mathmatica${}^{\text{\textregistered}}$ notebooks which can be found from the source at \url{https://arxiv.org/format/1901.10354}. The file \textsc{LargeC.nb} is concerned with explicit checks of the large $c$ eigenvalues of the second charge. This notebook uses \textsc{Virasoro.nb} by Matthew Headrick, which can be downloaded from \url{http://people.brandeis.edu/~headrick/Mathematica/Virasoro.nb}. Finally in \textsc{KdV\_charges.nb} we explicitly construct the leading order parts of the first $n$ charges in large $c$. It in particular confirms the coefficients we give in \S \ref{sec:higherK}. 

\begin{appendix}

\section{Classical integrals of motion}

\subsection{Integral of motions for the modified KdV equation} 

The classical integrals of motion defining the mKdV hierarchy are given by 
\begin{equation}
I_n = -\frac{1}{2\pi}\int_0^{2\pi} d\sigma \frac{1}{2}u(\sigma)Y_{2n-1}(u) \label{eq:mKdV}
\end{equation}
with 
\begin{equation}
Y_{n+1} = \partial_\sigma Y_n + u \sum_{k=1}^{n-1} Y_k Y_{n-k}\,,\quad Y_1 = -u\,.
\end{equation}

\subsection{Integral of motions for the KdV equation} 

The classical integrals of motion defining the KdV hierarchy are given by
\begin{equation}
K_n(v) = \frac{1}{2\pi} \int_0^{2\pi} d\sigma \frac12 v(\sigma) Z_{2n+1}(v)  \label{eq:KdV}
\end{equation}
with 
\begin{equation}
Z_{n+1} = \partial_\sigma Z_n + v \sum_{k=1}^{n-1}Z_k Z_{n-k}\,,\quad Z_1 =1\,.
\end{equation}

\section{Quantum KdV charge of $2d$ CFTs}

\subsection{Composite operators}

A composite operator $AB(z)$ of two quantum fields $A(z) = \sum\limits_n A_n z^{-n}$ and $B(z) = \sum\limits_n B_n z^{-n}$  is defined as 
\begin{equation}
 AB(z) \equiv \frac{1}{2\pi i} \oint\limits_{\mathcal{C}_z}\frac{d\zeta}{\zeta} \frac{z}{\zeta-z} \mathcal{R}(A(\zeta)B(z))
\end{equation}
with $\mathcal{R}$ denotes radial ordering, i.e. 
\begin{equation}
\mathcal{R}(A(\zeta)B(z)) = \left\{ \begin{matrix}
A(\zeta) B(z) \qquad& \text{for~} |\zeta|\ge |z|\\
 B(z)A(\zeta) \qquad& \text{for~} |\zeta|< |z|
\end{matrix}\right..
\end{equation}

\noindent
The modes of the composite field are then given by 
\begin{equation}
AB_n = \sum\limits_{l=1}^{\infty} A_{-l} B_{l+n} + \sum\limits_{l=0}^\infty B_{n-l}A_l\,.
\end{equation}

\noindent
This definition gives you some sort of normal ordering with respect to the modes $A_n$ and $B_n$. However, note that there is no additional normal ordering part. The composition of several operators is always defined by first composing the most left operators and successively composing the operators that follow to the right, e.g. $ABC(z)$ is defined by composing $AB(z)$ with $C(z)$. 

Furthermore, any composite operator from now on is understood as a symmetric composition, i.e. 
\begin{equation}
\begin{split}
 \left\langle AB\right\rangle = &\,\frac{1}{2} \left(AB(z) + BA(z)\right)\\
 \left\langle A_1A_2\dots A_n \right\rangle =&\, \frac{1}{n!} \sum_p A_{p_1}A_{p_2}\dots A_{p_n}
\end{split}
\end{equation}
where the sum is taken over all permutations and we omit the brackets in what follows.

\subsection{The charges}

The first few charges are given by (see also \cite{sy})
\begin{equation}\label{eq:CFTKdVs}
\begin{split}
  H \equiv \hat K_1 =&\, \frac{1}{2\pi\hbar' i}\oint \frac{dz}{z} \mathcal{L}(z) =L_0 - \frac{c-1}{24}\\
  Q \equiv\hat K_2 =&\,  \frac{1}{2\pi\hbar' i}\oint \frac{dz}{z}\,2 \mathcal{L}^2(z) =2\hbar' \left(2\sum\limits_{n=1}^\infty L_{-n}L_n+ L_0^2 -\frac{c-1}{12} L_0 + \frac{(1-c)^2}{576} \right)\\
  \hat K_3 =&\, \frac{1}{2\pi \hbar' i}\oint \frac{dz}{z}\, 8 \mathcal{L}^3 + \frac{2 \hbar'}{3}(c+2) (D\mathcal{L})^2\\
           =&\, 8 \hbar'^2 \Big( \sum\limits_{n,m=1}^\infty  2L_{-m}L_{-n+m} L_n + L_{-n-m}L_{m} L_n+L_{-n}L_{-m} L_{n+m} + \\
           & \quad +3 \sum\limits_{n=1}^\infty L_{-n} L_0 L_n-  \frac{2(c+2)n^2 -36 n +3 (c-1)}{12}L_{-n}L_n \\
           & \quad  + L_0^3 + \frac{1-c}{8} L_0^2 + \frac{(1-c)^2}{192} L_0 + \frac{(1-c)^3}{13824}\Big)\\ 
  K_4 =& \, \frac{40}{2\pi\hbar' i}\oint \frac{dz}{z}\, \Big[\mathcal{L}^4 + \frac{\hbar'}{3}(c+2) \mathcal{L}(D\mathcal{L})^2 + \\&+\frac{\hbar'^2}{180}(c+2)(c-\frac{1}{2})(D^2\mathcal{L})^2 - \frac{\hbar'^2}{24}(c+2)(D\mathcal{L})^2\Big]\,.
\end{split}
\end{equation}

\noindent
In the limit $\hbar'\to \frac{3}{c}\to 0$ they converge to the classical KdV charges \eqref{eq:KdV} when we replace $\mathcal{L} = \frac{v}{2}$ up to an overall factor of $\frac{1}{\hbar'}$. 

\subsection{Some explicit checks on large $c$ eigenvalues of Q}\label{b3}

We can in principle computed the eigenvalues of $\tilde Q^{(1)} = 2\sum_{n=1}^\infty L_{-n}L_n $ with the help of Mathematica up to arbitrary level. However, the calculation time grows rapidly with the level. We here give the results up to level 6 in the vacuum module and show that the large $c$ eigenvalues are given by $k_i \frac{n_i^3-n_i}{6} c$. 

\noindent
\textbf{Level 0 \& 1:\quad} $\tilde Q^{(1)}$ annihilates the vacuum and there is no level 1 in the vacuum module. So no contribution here.

\noindent\textbf{Level 2:\quad} A (not normalized) basis is $\{L_{-2}\ket{0}\}$. Its only basis element's eigenvalue is simply $c$ and matches $c\frac{n^3-n}{6} \stackrel{n=2}{=} c$.

\noindent\textbf{Level 3:\quad} A (not normalized) basis is $\{L_{-3}\ket{0}\}$. Its only basis element's eigenvalue is $4 c + 4$. In the large $c$ limit this gives $4 c = \frac{3^3-3}{6} c$. 

\noindent\textbf{Level 4:\quad} A (not orthonormal) basis is $\{L_{-4}\ket{0},L_{-2}^2\ket{0}\}$. The matrix representation of $\tilde Q^{(1)}$ in this basis is
\begin{equation}
\begin{pmatrix}
10 c +20 & 3c +6\\
6 & c+8
\end{pmatrix}
\end{equation} 
with eigenvalues $\lambda_{1/2} = 9 + 3c \pm \sqrt{37+22 c + 4 c^2}$. At large $c$ this gives 
\begin{align}
\lambda_1 &= 10 c = \frac{4^3-4}{6} c\\ 
\lambda_2 &= 2 c = 2 \frac{2^3-2}{6} c 
\end{align}

\noindent\textbf{Level 5:\quad} A (not orthonormal) basis is $\{L_{-5}\ket{0},L_{-3}L_{-2}\ket{0}\}$. The matrix representation of $\tilde Q^{(1)}$ in this basis is
\begin{equation}
\begin{pmatrix}
20 c +50 & 9c +18\\
30 & 5c +50
\end{pmatrix}
\end{equation} 
with eigenvalues $\lambda_{1/2} = \frac{100 + 25c \pm \sqrt{45\left(48+24 c + 5 c^2\right)}}{2}$. At large $c$ this gives 
\begin{align}
\lambda_1 &= 20 c = \frac{5^3-5}{6} c \\
\lambda_2 &= 5 c =  \left(\frac{3^3-3}{6} + \frac{2^3-2}{6}\right) c 
\end{align}

\noindent\textbf{Level 6:\quad} A (not orthonormal) basis is $\{L_{-6}\ket{0},L_{-4}L_{-2}\ket{0},L_{-3}L_{-3}\ket{0},L_{-2}L_{-2}L_{-2}\ket{0}\}$. The matrix representation of $\tilde Q^{(1)}$ in this basis is
\begin{equation}
\begin{pmatrix}
35 c + 88&12c&36c+72&48c+48\\
36&11c+42&108&18c+132\\
18&24&8c+52&48\\
0&11&0&3c+48
\end{pmatrix}\,.
\end{equation} 

\noindent At large $c$ the four eigenvalues are
\begin{align}
\lambda_1 &= 35 c = \frac{6^3-6}{6} c\\
\lambda_2 &= 11 c = \left(\frac{4^3-4}{6}+\frac{2^3-2}{6}\right) c\\
\lambda_3 &= 8 c = 2\frac{3^3-3}{6} c\\
\lambda_4 &= 3 c = 3 \frac{2^3-2}{6} c
\end{align}

\subsection{Global Q-character}

By the global character we denote the object that follows from taking the trace over states in the sub-module corresponding to the global sub-algebra of the full Virasoro algebra, i.e. $sl_2 \subset \text{Vir}_c$. The $sl_2$ sub-module $[p]\big|_{sl_2}$ of the primary module $[p]$ is built upon states of the form $L_{-1}^k\ket{p}$, $k\in\mathbb{N}$.    

We also have to restrict $Q$ to the parts that act within this global sub-module. In the present case this means we only keep the terms that include $sl_2$ generators,
\begin{equation}
Q_{sl_2} \equiv 2\hbar'\left(2 c \, L'_{-1}L'_1 + {L_0}^2 -\frac{c-1}{12}L_0 \right) \equiv Q_{sl_2}^{(1)} + 2\hbar' L_0^2 - 2\hbar'\frac{c-1}{12} L_0\,.
\end{equation}

\noindent 
All the terms that we omitted act like
\begin{align}
\begin{split}
2\hbar' c L'_{-m}L'_m {L'}_{-1}^k \ket{p} &= 2\hbar'\frac{(m+1)!}{c^{\frac{m-1}{2}}}\left(\begin{matrix}k\\k-m\end{matrix}\right)\, \left( \frac{k-m}{m+1}  +  p\right) {L'}_{-m}{L'}_{-1}^{k-m} \ket{p}\\
&\equiv Q_{k,m,p}\,  L'_{-m}{L'}_{-1}^{k-m} \ket{p} \notin [p]\big|_{sl_2}\quad \text{for} ~m>1\,.
\end{split}
\end{align}

\noindent
They do not act within the global sub-representation and hence have to be omitted. 

The $Q_{sl_2}^{(1)}$ eigenvalues of a global state $L_{-1}^k \ket{p}$ are exactly the one in \eqref{eq:sl2EV} with $h_\psi = p$, so
\begin{equation}
Q_{sl_2}^{(1)} L_{-1}^k \ket{p} = 4 \hbar'\left(k^2 +(2p-1)k \right)L_{-1}^k \ket{p}\,.
\end{equation}

\noindent
It follows immediately that the $Q_{sl_2}^{(1)}$ character in the global module $[p]\big|_{sl_2}$ is given by
\begin{equation}\label{eq:globalQ1}
 \Tr_{[p]|_{sl_2}} y_1^{Q_{sl_2}^{(1)}} = \left(\sum_{k=0}^\infty y_1^{4 \hbar'\left(k^2 +(2p-1)k \right)}\right)\,.
\end{equation}

\noindent
The $L_0$ eigenvalues of the previous states are simply $(p+k)$. The full $Q_{sl_2}$ character is, hence, given by
\begin{equation} \label{eq:FullQsl2}
\Tr_{[p]|_{sl_2}} q^{Q_{sl_2}} = \sum_{k=0}^\infty y_1^{2\hbar'\left(2k^2 +(2p-1)2k + (p+k)^2 - \frac{c-1}{12}(p+k) \right) }\,.
\end{equation} 

\noindent
We see here that we cannot simply interchange the limit and the summation. This is because the eigenvalues switch sign at $k\sim c$ which completely changes the behavior of the sum. In particular, when $c$ is very large, then $k\lesssim c$ gives huge contributions to the character!  

\subsubsection{Closed forms for Global Q-characters}
\label{Q-closed}

The eigenvalues $$\lambda(k) = 2\hbar'\left(2k^2 +(2p-1)2k + (p+k)^2 - \frac{c-1}{12}(p+k) \right)$$ are negative between $$k_\pm =\frac{c+23-72p\pm\sqrt{529+46c+c^2-3456p+3456p^2}}{72}\,.$$ 

\noindent
Hence, for large $c\gg p$, $\lambda$ is negative between $k_-\!<\!k_0=0$ and $k_+ \approx \frac{c}{36}$. The minimum lies at $k_\text{min} = \frac{23+c-72p}{72} \approx \frac{c}{72}$, with a minimal value of $$\lambda(k_\text{min}) = -\hbar\frac{c^2+46c+3456p^2-3456p+529}{864}\approx -\frac{\hbar c^2}{864}\,.$$

\noindent
The sum \eqref{primary-sum} for $k< k_+$ can be approximated by an integral for $\mu\gg 1/c$ and $c\gg p$. It gives 
\begin{equation}
 \sum_{k=0}^{k_+} y_1^{\lambda(k)} \approx\sqrt{\frac{\pi c}{18\mu_1}} \exp\left[\mu_1\left(\frac{c}{288} +\frac{23}{144}+ \frac{12 p(p-1)+\frac{529}{288}}{c}\right)\right]\,. 
\end{equation}

\noindent
It is obvious that the contribution to the global character by negative eigenvalues is huge. 

Note, that in case of higher charges one also needs to consider the first sub-leading contribution from the charge. Higher order corrections in $1/c$ are not needed to render the sum finite. The global character looks like \eqref{eq:FullQsl2} but dressed with coefficients that follow from \eqref{Kk}. 

Note further, that it seems reasonable that the global character behave somehow nice under modular transformation, since their form looks much like the well known $\theta$ functions. If we for example take the primary to be $p=\frac{1}{2}$, then \eqref{eq:globalQ1} is given by $\frac12\left(\theta_3(\tau)+1\right)$ with $\tau = i\frac{24 \mu_1}{c \pi}$ and 
$$\theta_3(\tau) = \sum_{n\in\mathbb{Z}} q^{\frac{n^2}{2}}\,.$$

\noindent
Or for $p=1$ it can be written as $\frac{q^{1/4}}{2} \theta_2(\tau)$ with 
$$\theta_2(\tau) = \sum_{n\in\mathbb{Z}} q^{\frac{\left(n+\frac12\right)^2}{2}}\,.$$

\noindent
It is possible to write the general global character in any primary module as the generalized $\theta$-function given by 
\begin{equation}
 \begin{split}
\theta \!\left[\alpha,\beta\right]\!(\tau,z) &= \sum_{n\in \mathbb{Z} } q^{\frac{1}{2}(n+\alpha)^2}\, e^{2\pi i (n+\alpha)(z+\beta)} \\
 &= \frac{\eta(\tau) \,e^{2 \pi i \alpha (z+\beta)}}{q^{\frac{\frac{1}{24}-\alpha^2}{2}}}\prod _{n=1}^\infty \left(1\!+\!q^{n+\alpha-\frac{1}{2}}\,e^{2\pi i (z+\beta)}\right)\! \left(1\!+\!q^{n-\alpha-\frac{1}{2}}\,e^{-2\pi i(z+\beta)}\right)\,,
 \end{split}\label{eq:genTheta}
\end{equation}
which behave under modular transformations as
\begin{equation}
\begin{split}
\theta[\alpha,\beta](\tau+1,z) &= e^{-i\pi\alpha(\alpha-1)}\theta[\alpha,\alpha+\beta-\frac{1}{2}](\tau,z)\,,\\
\theta[\alpha,\beta](-\frac{1}{\tau},\frac{z}{\tau}) &= \sqrt{-i\tau}e^{2\pi i\alpha\beta+i\pi \frac{z^2}{\tau}}\theta[\beta,-\alpha](\tau,z)\,.
\end{split}
\end{equation}

\section{Saddle}
We want to evaluate by saddle the following integral : 
$$
\int dn \exp\left( A(l) n^{1/2l} - \mu_l n \right). 
$$
The saddle point equation yields,  
\bea
\frac{A(l)}{2l} n_s^{\frac{1-2l}{2l} } -\mu_l &=& 0. \nn \\
\text{or, \qquad} n_s &=& \left( \frac{2l \mu_l}{A(l)} \right)^{\frac{2l}{1-2l} }.
\eea
Plugging back the saddle point into the integral we obtain,  
\bea
\exp \left( A(l) n_s^{1/2l} \right) y_l^{n_s} &=& 
\exp \left( A(l) \left(\frac{2l \mu_l}{A(l)}\right)^{\frac{1}{1-2l} } \right) y_l^{ (\frac{2l \mu_l}{A(l)} )^{\frac{2l}{1-2l} } }, \nn \\
&=& \exp \left( A(l)^{\frac{2l}{2l-1}} (2l \mu_l)^{\frac{1}{1-2l} } (  1  - (2l)^{-1} )\right) 
\eea
Thus,
\be
\int dn \exp\left( A(l) n^{1/2l} \right) y_l^n \approx
\exp \left\{  \left(  1  - \frac{1}{2l} \right) A(l)^{\frac{2l}{2l-1}} \left(2l \mu_l\right)^{\frac{1}{1-2l} }\right\}.
\label{saddle1}\ee

\bibliographystyle{JHEP}  
\bibliography{refs}

\end{appendix}




\end{document}